# Comment on "Reconciling results of tunnelling experiments on (Ga,Mn)As" arXiv:1102.3267v2 by Dietl and Sztenkiel

Shinobu Ohya, Kenta Takata, Iriya Muneta, Pham Nam Hai, and Masaaki Tanaka

*Department of Electrical Engineering and Information Systems, The University of Tokyo,*

*7-3-1 Hongo Bunkyo-ku, Tokyo 113-8656, Japan.*

Abstract:

We comment on the recent paper "Reconciling results of tunnelling experiments on (Ga,Mn)As" arXiv:1102.3267v2 by Dietl and Sztenkiel.[1] They claimed that the oscillations observed in the $d^2I/dV^2$-$V$ characteristics in our studies on the resonant tunneling spectroscopy on GaMnAs[2,3] are not attributed to the resonant levels in the GaMnAs layer but to the two-dimensional interfacial subbands in the GaAs:Be layer. Here, we show that the interpretation shown in Ref. 1 is not able to explain our experimental results and our conclusions[2,3] remain unchanged.

First, we show that the scenario of Ref. 1 contradicts the GaMnAs thickness ($d$) dependence of the resonant peaks shown in our papers[3,4] [for example, Fig. 3**a**,**b** in Ref. 3 and Fig. 1(d) in Ref. 4]. Figures 1 (a) and (b) show the resistance-area product ($RA$) as a function of $d$ of the $Ga_{1-x}Mn_xAs$($d$ nm)/ AlAs(5 nm)/ GaAs:Be diode device with the Mn concentration $x$ of 6% and the Curie temperature of 71 K (named as Sample A in Ref. 3) when the bias voltage $V$ is -10 and -100 mV, respectively. The $RA$ changes are only ~30% both in (a) and (b) in the wide range of $d$ except for the small $d$ region ($d$ < 3 nm). The increase of $RA$ in the small $d$ region is due to the surface depletion in GaMnAs caused by the Fermi level ($E_F$) pinning at the surface and the thin band bending at the interface between GaMnAs and AlAs. (See Fig. 2**a** and **b** in Ref. 3.) The results in Fig. 1 (a) and (b) show that the $RA$ change is small in the wide range of $d$ at both bias voltages of -10 mV and -100 mV when $d$ > 3 nm. This means that the band profiles near the interfaces of GaMnAs/AlAs and AlAs/GaAs:Be, thus the subband levels formed at the interface between AlAs and GaAs:Be, are not largely modified by changing $d$, when $d$ > 3 nm.

In the scenario of Ref. 1, the resonant peak bias voltage is proportional to $RA(d)$. Thus, the $d$ dependence of the resonant peak bias voltage is quite similar to the $RA$-$d$ curve. This can be seen by comparing the calculated results shown in Fig. 1**c** of Ref. 1 and the experimentally obtained $RA$ values shown in Fig. 1(b). In our resonant tunneling experiments, however, the shifts of the resonant peaks with increasing $d$ are clearly different depending on the resonant levels. For example, $RA$ of Sample A changes from 0.0549 to 0.0389 $\Omega\text{cm}^2$, which corresponds to a change of 29% (=1-0.0389/0.0549), when $d$ is changed from 5.0 to 17.4 nm at $V$= -100 mV as shown in Fig. 1(b). As can be seen in Fig. 5**a** of Ref. 3, however, the shifts of the resonant peaks with increasing $d$ from 5 to 17.4 nm are 12% (-130 → -115 mV), 28% (-180 → -130 mV), and 41% (-252 → -148 mV) for HH1, LH1 (HH2), and HH3, respectively. It is clear that the shifts of HH1 (12%) and HH3 (42%) are largely different from the change of *RA* (~30%). Therefore, this experimental result contradicts the scenario in Ref. 1.

In Fig. 1c of Ref. 1, the fitting to the first level is worse than that to the second level. This is because the shift of the first level (HH1) with increasing *d* (=12%) is smaller than that of the second level (LH1 and HH2) (=28%) which coincidentally has a similar change of *RA* (=29%) with increasing *d*. Moreover, as expected from the above consideration, the fitting to the higher levels will be worse because the higher levels more largely move than *RA* with increasing *d* as shown in Fig. 5a of Ref. 3. This fundamental difficulty of the fitting to our experimental results in Ref. 1 means that the origin of the $d^2I/dV^2$ oscillation observed in our study is *not* due to the two-dimensional interfacial subbands in the GaAs:Be layer.

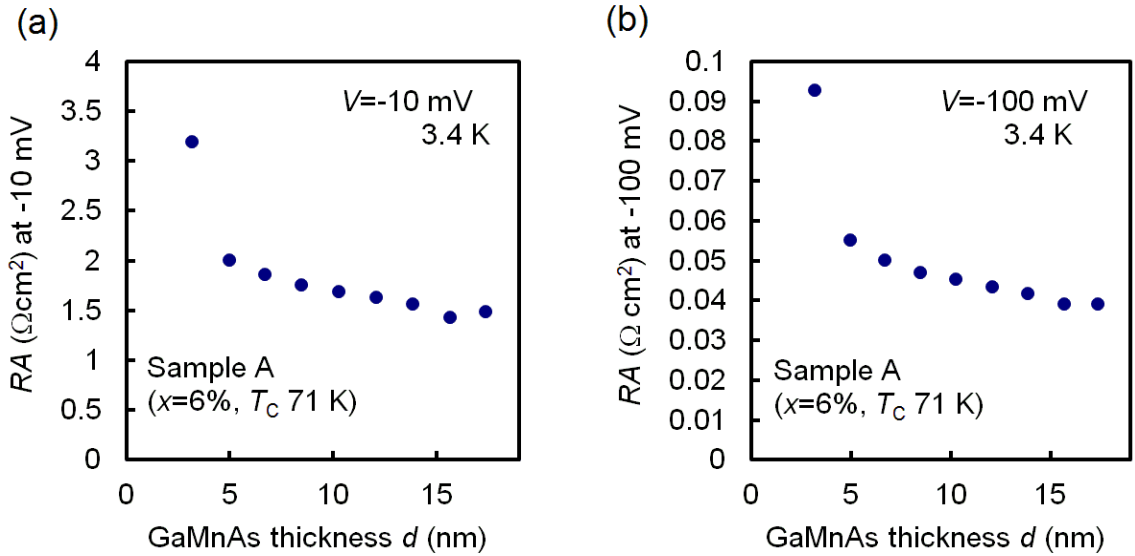

Figure 1  Resistance area product (*RA*) as a function of GaMnAs thickness *d* at (a) *V*= -10 mV and (b) -100 mV in our diode device containing $Ga_{0.94}Mn_{0.06}As$ with the Curie temperature of 71 K (named as Sample A in Ref. 3.)

The essential result obtained in our resonant tunneling spectroscopy is that the VB gradually branched into subband levels with decreasing *d* due to the quantization of VB of GaMnAs. Our experimental results clearly show this typical behavior of the quantization. For example, in Fig. 5a and b in Ref. 3, LH1 (HH2) level looks as if it branched from the HH1 level at *d*=~15.7 and ~14.6 nm, respectively. Also, it is clear that other resonant peaks tend to be merged with the HH1 level at a certain *d* (*d*>20 nm) with increasing *d* in Fig. 5a and b in Ref. 3. Therefore, it is quite natural to conclude that these peaks are formed by the quantization of VB in GaMnAs. The scenario in Ref. 1 cannot explain this essential feature of the resonant peaks obtained in our study, because the band profile of the interface AlAs/GaAs:Be is not largely modified and thus the two-dimensional interfacial subbands at GaAs:Be are never merged by increasing $d$.

Another problem of the scenario of Ref. 1 is that it cannot explain our experimental results of the weakening of the resonant peaks with increasing $d$ (Fig. 3**a**,**b** in Ref. 3). This phenomenon is a typical feature of the resonant tunneling effect when the thickness of the quantum well is increased. If the resonant levels formed at the AlAs/GaAs:Be interface appeared in the *I-V* characteristics, such an intensity weakening would not be observed, because the band profile of this interface is hardly modified when $d > 3$ nm. Therefore, the model of Ref. 1 contradicts our experimental results.

Although the two-dimensional interfacial subbands formed at the AlAs/GaAs:Be interface have been reported when the Be concentration is extremely low (for example, $6\times10^{14}$ cm$^{-3}$ in Ref. 5) as indicated in Ref. 1, it seems to be difficult to detect these states when the Be concentration is as high as that used in our study ($\sim 10^{18}$ cm$^{-3}$). We do not see any clear signals related to the interfacial states at AlAs/(GaAs:Be or a thin GaAs spacer) in the studies of the similar p-type GaAs-based resonant tunneling structures[6,7,8], which is probably due to the very weak quantization of holes with the quantization energies of only less than 10 meV at this interface, as also shown in Fig. 1**b** of Ref. 1.

To confirm this fact, we fabricated Ga$_{0.947}$Mn$_{0.053}$As(100 nm)/ AlAs(5 nm)/ GaAs:Be (Be: $1\times10^{18}$ cm$^{-3}$, 100 nm) on a p$^+$GaAs(001) substrate. We annealed this sample at 160˚C for 20 hours. By measuring the magnetic circular dichroism on this annealed sample, $T_C$ was estimated to be 75 K. We reduced the thickness of the surface GaMnAs layer from 100 nm to 30 nm by etching and fabricated a tunneling diode device with the same method used in Ref. 3. The $d^2I/dV^2$-$V$ curve obtained in this device (named as Sample R) at 3.5 K is shown in Fig. 2(a). A few peaks were clearly observed in the negative bias region. Here, we divide the negative bias region into the three regions named as (b), (c), and (d) as described in Fig. 2(a). Corresponding schematic band diagrams are shown in Fig. 2(b), (c), and (d), respectively.

The peak X observed at ~ -30 mV in region (b) was also observed in other GaMnAs tunneling devices as shown in Fig. 6**b** and **c** in Ref. 3. This peak is not largely moved as a function of $d$, and it is enhanced by increasing the Mn concentration or $T_C$. Therefore, this peak is probably a part of the impurity states (or band) in the bandgap. The $d^2I/dV^2$ change in region (c) corresponds to the VB edge of GaMnAs. This VB edge position is almost the same as that of Sample A (~ -120 mV) as shown in Fig. 5a in Ref. 3, which means that the bias dependence of the band deformation in Sample R is quite similar to that of Sample A in Ref. 3. Region (d) corresponds to the VB region in GaMnAs. In our previous studies on the tunneling diodes with the GaMnAs thickness less than 20 nm, oscillations due to the quantization of the VB of GaMnAs were seen in region (d) as shown in Fig. 2**a-c** and Fig. 5**a-c** of Ref. 3. However, these oscillations were absent in Sample R where the GaMnAs thickness is 30 nm. This result is because the quantization of VB is quite weak in Sample R. If two-dimensional interfacial states at AlAs/GaAs:Be were observed in our device, they would appear in region (d) in Fig. 2(a) regardless of the GaMnAs thickness because the triangular shaped VB profile of GaAs:Be is formed only in region (d) as shown in Fig. 2(d). However, there is no clear peak in region (d) in Sample R, which means that the interfacial states are not observed in our study at all.

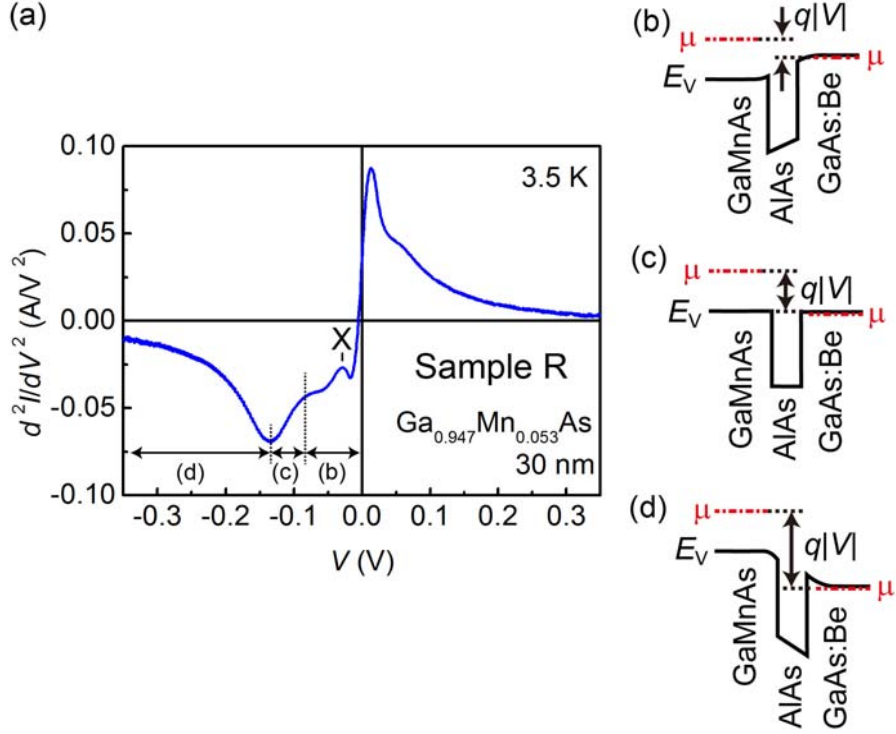

Figure 2 (a) $d^2I/dV^2$-$V$ curve obtained in Sample R composed of GaMnAs(30 nm)/ AlAs(5 nm)/ GaAs:Be(Be: 1×10$^{18}$ cm$^{-3}$, 100 nm) on a p$^+$GaAs(001) substrate at 3.5 K. (b), (c), (d) Schematic band diagrams corresponding to region (b), (c), and (d) defined in (a), respectively. The black curves $E_V$ are the VB edge, and the red dotted lines μ are the chemical potentials of the electrodes. Here $q$ is the elementary charge.

In conclusion, the $d^2I/dV^2$ oscillations observed in our experiments[3,4] are caused by the resonant tunneling effect in the GaMnAs layer, not by that at the interface between the AlAs and GaAs:Be layers.

Finally, Ref. 1 claimed that our resonant tunneling experiments contradict the recent results of the scanning tunneling microscopy (STM) reported by Richardella *et al.*[9] As already explained in Ref. 3, we think that our results do not contradict the STM results. In our opinion, the STM results seem to be consistent with, and at least not contradictory to, the impurity-band (or state) conduction model.